# Superchords: the atoms of thought[1]

NORMAND, Rogério & FERREIRA, Hugo Alexandre
Institute of Biophysics and Biomedical Engineering (IBEB)
Faculty of Science, University of Lisbon

**Abstract**

Electroencephalography (EEG) signals' interpretation is based on waveform analysis, where meaningful information should emerge from a plethora of data. Nonetheless, the continuous increase in computational power and the development of new data processing algorithms in the recent years have put into reach the possibility of analysing raw EEG signals. Bearing that motivation, the authors propose a new approach using raw data EEG signals and deep learning neural networks, for the classification of motor activities (executed and imagery). The hypothesis to be presented here is: each instantaneous measurement of the raw signal of all EEG channels (superchord) is unique per motor activity regardless the moment of measurement. This study has confirmed the hypothesis (results with accuracy over 80%, mean for 109 subjects), reinforcing the need of further research for the understanding of mental processes.

**Subject Terms:** Artificial Neural Networks, Brainprint, Deep Learning, EEG, Flatcharts, Neuroscience, Superchords.

**Introduction**

Electroencephalography (EEG) records the scalp voltage of brain activities with a high temporal resolution. Microstates are defined as "quasi-stable spatial distributions (landscapes) of electric potential that are connected by quick changes in landscapes" (Lehmann et al, 2009). Our brains present a quasi-stable brain topography for sub-second periods of time intercalated with abrupt changes. Those processes are believed to be connected with brain functional states during the execution of activities and are generated by changes on the brain electric field.

The authors would like to propose a new approach potentially useful for better understanding of mental processes through EEG. In the Brain Orchestra Approach[2] (BOA), a superchord is defined as the instantaneous measurement of the raw signal of all EEG channels, analog to a musical chord[3].

The hypothesis tested here is that for any single motor activity, each superchord is unique and it can be used to determine that activity, regardless the moment of measurement.

---

[1] A small homage to Dr. Dietrich Lehmann (Dec. 3, 1929 - Jun. 16, 2014) - Lehmann, 1990.

[2] One neurone is equivalent to a note; a group of neurones equals to an instrument; a group of similar instruments is like a brain region; so, the brain is like an orchestra. Electrodes work like microphones, capturing the electric signals from brain processes with "louder" signals from regions close to them.

[3] Superchords are represented in a straight horizontal line of pixels, where each one represents a channel and the colour scale varies from bright green to bright red, respectively for higher and lower than mean raw voltage values. Black colour represents the mean of all channels in the superchord.

In order to facilitate visualisation, authors have developed the concept of a flatchart[4], where it is possible to identify the brain signature/pattern for a specific task, the activity brainprint[5]. Fig. 1 pictures brainprints of making-a-fist activity with left and right hand.

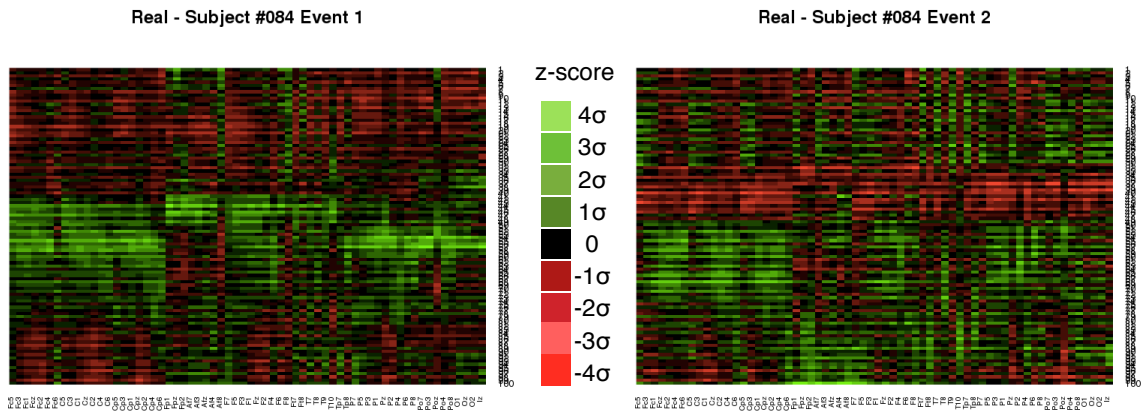

Figure 1: On the left, the brainprint of a left hand make-a-fist, displaying 100 superchords (y-axis) and 64 channels (x-axis), for subject 84, where red/green scale represents negative/positive values in terms of standard deviations for each channel. On the right side, the same for the right hand. The period of each superchord is 6.25 ms (1/160 Hz).

**Methodology**

Using a public dataset (Goldberger, 2000; Schalk, 2004), R language (R Core Team, 2014; RStudio Team, 2012) and H2O[6] Deep Learning algorithms (Ambati, Candel, Click et al., 2015), it was possible to test the hypothesis.

The dataset was produced from a visual stimulus/motor response paradigm with 5 possible reactions (movements with left hand, right hand, both hands, both feet or resting), also segregated into real actions or imagery (R/I), for 109 subjects with a sampling rate of 160 Hz (period = 1/160 = 6.25 ms).

The methodology applied was to read the datasets for each subject, adding proper label for each performed task, having columns as channels/task label and measurements per period for each variable (channels/task label) in rows (superchords). Considering that the mind focus is higher just after the stimulus, the complete dataset is sliced to consider only the first 10, 20, 30, 40, 60, 80, 100, 120, 140 and 160 first superchords (time range from 62.5 to 1000 ms)[7] for each stimulus, originating ten new sliced datasets.

Each subject sliced dataset is divided in two equal parts after a H2O built-in shuffle per row (randomisation), being the first part used for training and the second used for validation purposes.

---

[4] A flatchart is the result of chronologically juxtaposing superchords, representing in a 2D chart a 4D process (3D spatial one for the brain and one for time). Through flatcharts it is possible to observe the whole event at once. Flatcharts have superchords in rows and channels in columns.

[5] A brainprint is the specific signature of a brain activity displayed in a flatchart. It is built through superimpositions of same activity flatcharts, previously normalised by channel (mean equals to 0 and divided by standard deviation). The final flatchart is normalised once more to be independent of the number of superimposed flatcharts. Red/Green colour scale is used to emphasise lower/higher values (voltage).

[6] H2O is the leading open source machine learning project from the H2O.ai community.

[7] Each dataset was composed by 66 columns and a different number of superchords/rows (according to time slices). The first column (timestamp) was not used. Columns 2 to 65 represent each channel raw data and column 66 is the label for one of the 5 possible tasks.

A programming script in R/RStudio[8] was developed to feed the training dataset into the H2O classification algorithm in order to generate a model to be tested against the validation dataset. The classification error was calculated as the number of wrong predictions divided by the number of total predictions and the mean error was defined as the sum of each case subject error divided by the total number of subjects (109).

**Results & Discussion**

In the fig. 2 below it is possible to compare the worse and the best case scenarios for this study.

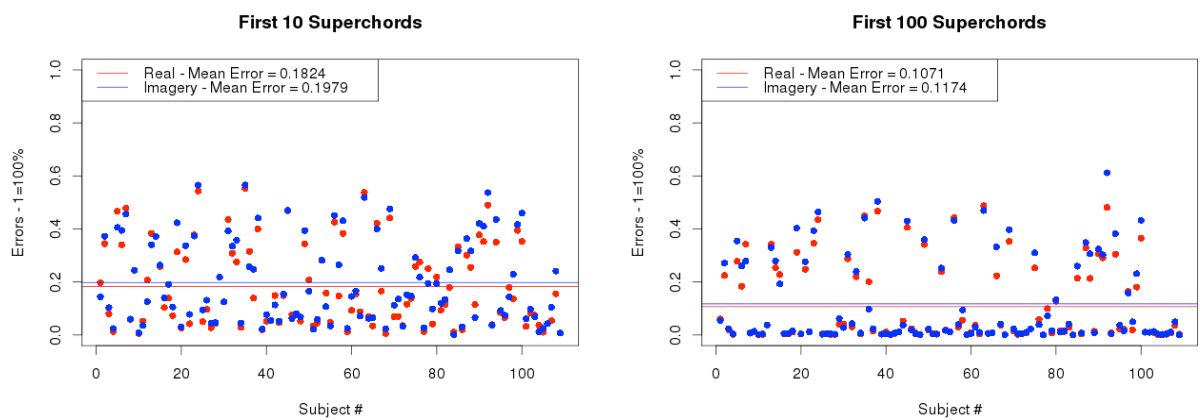

Figure 2: On the left, the error measurement for each subject is shown considering only the first 10 superchords after the stimulus, being red for the real actions and blue for the imagery. Horizontal lines correspond to the mean error for each set (Real ≈ 18.24%; Imagery ≈ 19.79%). On the right side, the error results for the same subject are shown, but now considering the first 100 superchords (Real ≈ 10.71%; Imagery ≈ 11.74%).

In the worse case, with only 10 superchords after the stimulus has been presented, the results have a mean error of ≈18.24% for real movements and ≈19.79% for imagery. By itself, this is a strong encouragement towards the correctness of the hypothesis. The best case, with 100 superchords, the errors, on average, were ≈10.71% and ≈11.74%, respectively for real/imagery datasets, another strong evidence to support the correctness of the proposed hypothesis. A chart of the mean error per superchord time slice is displayed in fig. 3 - left side.

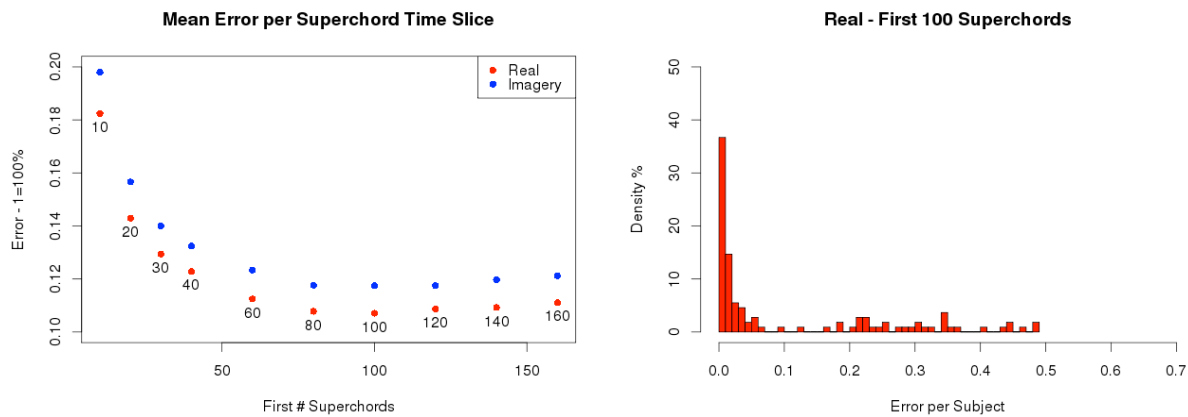

Figure 3: On the left, the mean error for each superchord time slice (10, 20, 30, 40, 60, 80, 100, 120, 140 and 160) is shown. Y-axis is defined between 0.10-0.20 to better display the differences among the results. On the right side, the error per subject for the first 100 superchords is shown in a density histogram measured in percentage and for real actions.

---

[8] R scripts could be provided on a case-by-case basis. Please contact the authors.

It is noteworthy to mention that, with 100 superchords datasets for real actions, 39 subjects presented errors below 1% (69 lower than 5%, 74 lower than 10%), meaning that for those individuals, with a single superchord, regardless the moment of measurement, it would be possible to determine their task/mental activity in each of the 5 possibilities with great precision. In fig. 3 - right side shows the density distribution of errors for real actions and first 100 superchords (≈67% of subjects with less than 7% error)[9].

Nevertheless, some considerations deserve attention:

1. Based on the paradigm protocol, the experiment was not totally random as each action task should be followed by a resting task. At each round, only 3 task options were available for each subject, no matter if real or imagery. Due to that, there is the possibility of subject anticipation (Chavarriaga, 2012), which allowed an individual to be ready for the next task prior to the stimulus;
2. The quality of the measurements, despite the best efforts employed: there is also a considerable chance of electrode misplacement, saline bridges, low signal-to-noise ratio, etc.;
3. Sample representativity: there is no information regarding how subjects were selected and if they are representative of the population, as well as to their mental health, use of drugs or any other factor that may have contributed to measurement differences;
4. Paradigm type: the protocol involved only real or imagery motor reactions to stimuli.

On the other hand, if 1) is true, it would reinforce the proposed hypothesis, as anticipation would create specific superchords for a mental task. The others aspects should be objectively considered after new tests with specific protocols to avoid possible biases. Although the dataset referred motor activities, the time slices considered, as well the presence of imagery rounds, points towards the possibility of the correctness of the hypothesis for other mental activities.

## Conclusions

Despite more work is required on this subject, there are positive indications that superchords are singular to each motor task present on the dataset. The authors do believe it would be possible to expand the concept of singular superchord to other mental activities, like reactions to faces or words (another paper, covering this possibility, is being prepared).

More than to provide definitive conclusions, the authors would like to collaborate with those interested to apply this methodology to other paradigms, experiments and cases in order to promote advancements in neuroscience.

## Acknowledgements

The authors thank Fundação para a Ciência e Tecnologia and Ministério da Ciência e Educação Portugal (PIDDAC) for financial support under the grant UID/BIO/00645/2013. Authors also would like to convey their thanks to IBEB/FCUL, the Portuguese National Distributed Computing Infrastructure and H2O team for valuable contributions and support during all phases of the study.

---

[9] The hardware employed to obtain the results was a cluster of 16 CPUs with 64 GB RAM running on Linux. The use of lower amount of RAM can reduce the accuracy in up to 1 percent point. Due to the nature of Deep Learning neural networks, the results can present slight variations (up to ±0.5 percent point).